\documentclass[aps,prl,twocolumn,superscriptaddress,showpacks]{revtex4-2}

\usepackage{comment}
\usepackage[colorlinks, citecolor=red]{hyperref}
\usepackage{hyperref}
\usepackage[utf8]{inputenc}
\usepackage[T1]{fontenc}    %
\usepackage[english]{babel}
\usepackage[pdftex]{graphicx}
\usepackage{amsmath,amssymb,amsthm,bbm, mathtools} %
\usepackage{mathrsfs}
\usepackage{tikz}   %
\usepackage[normalem]{ulem}

\renewcommand{\vec}[1]{\boldsymbol{#1}}

\begin{document}

\title{Theory of quantum decoherence and its application to anomalous Hall effect}

\author{Xian-Peng Zhang}

\affiliation{Centre for Quantum Physics, Key Laboratory of Advanced Optoelectronic Quantum Architecture and Measurement (MOE), School of Physics, Beijing Institute of Technology, Beijing, 100081, China}

\affiliation{International Center for Quantum Materials, Beijing Institute of Technology, Zhuhai, 519000, China}

\author{Yan-Qing Feng}
\email{yq_feng@bitzh.edu.cn}

\affiliation{International Center for Quantum Materials, Beijing Institute of Technology, Zhuhai, 519000, China}

\author{Haiwen Liu}
\affiliation{Center for Advanced Quantum Studies, School of Physics and Astronomy, Beijing Normal University, Beijing 100875, China}

\author{Wanxiang Feng}
\email{wxfeng@bit.edu.cn}
\affiliation{Centre for Quantum Physics, Key Laboratory of Advanced Optoelectronic Quantum Architecture and Measurement (MOE), School of Physics, Beijing Institute of Technology, Beijing, 100081, China}
\affiliation{International Center for Quantum Materials, Beijing Institute of Technology, Zhuhai, 519000, China}

\author{Yugui Yao}
\email{ygyao@bit.edu.cn}
\affiliation{Centre for Quantum Physics, Key Laboratory of Advanced Optoelectronic Quantum Architecture and Measurement (MOE), School of Physics, Beijing Institute of Technology, Beijing, 100081, China}

\affiliation{International Center for Quantum Materials, Beijing Institute of Technology, Zhuhai, 519000, China}

\begin{abstract}
Coherent quantum phenomena can only emerge when decoherence is minimized, and mastery over decoherence is technologically crucial for designing and operating functional quantum devices. However, its microscopic mechanisms in spin–orbit–coupled ferromagnets remain elusive, and quantitative treatments have long been challenging. To solve this fundamentally significant and technologically crucial problem, we develop a quantum master-equation framework with a general ansatz for the off-diagonal density matrix that simultaneously captures electric-field–driven coherence and impurity-scattering–induced decoherence. This unified approach enables quantitative analysis of how decoherence reshapes the intrinsic anomalous Hall effect, revealing a clear crossover between intrinsic and extrinsic regimes. Remarkably, we identify a previously unrecognized extrinsic contribution: a \textit{second-order} scattering process tightly relative to quantum decoherence—that is fundamentally distinct from both skew scattering and side jump mechanisms, yet substantially more significant than the skew scattering mechanism. Our work establishes decoherence as a key element in quantum transport and provides a systematic extension of the Boltzmann transport equation to incorporate decoherence, with broad implications for robust spintronic functionality.
\end{abstract}

\maketitle

\textit{Introduction–} Quantum coherence, a cornerstone of quantum mechanics, occurs when a particle (e.g., an electron) simultaneously occupies multiple quantum states~\cite{zurek2003decoherence,streltsov2017colloquium}:
\begin{align} \label{fvdfvaml}
|\psi(t)\rangle = \sum_{\alpha} c_{\alpha}(t)|\alpha\rangle,
\end{align}
with $c_{\alpha}(t)=c_{\alpha}e^{-i\epsilon_{\alpha}t/\hbar}$ and $|\alpha\rangle$ denoting the eigenstates of the unperturbed Hamiltonian, e.g., $|\alpha\rangle=|\vec{k}\eta\rangle$.
The density matrix $\varrho(t)=|\psi(t)\rangle\langle\psi(t)|$ provides an alternative description: the diagonal terms $\varrho_{\alpha\alpha}$ give state populations, while the off-diagonal ones $\varrho_{\alpha\alpha’}$ ($\alpha\neq\alpha’$) encode quantum coherence~\cite{zhang2024microscopic}. Such coherence allows quantum interference, underlying quantum computation~\cite{nielsen2010quantum,breuer2002theory,deutsch1992rapid,grover1996fast} and transport~\cite{sinova2015spin,chang2023colloquium}.
When coherence persists in solids, it leads to striking transport phenomena, including the intrinsic anomalous Hall effect (AHE)~\cite{sundaram1999wave,jungwirth2002anomalous,culcer2003anomalous,fang2003anomalous,yao2004first} and spin Hall effect~\cite{sinova2004universal,murakami2003dissipationless,guo2008intrinsic,yao2005sign}, as well as their quantized counterparts~\cite{chang2023colloquium,chang2013experimental,nagaosa2010anomalous,bernevig2006quantumexp,liu2008quantum,wang2014quantum,kane2005quantum,bernevig2006quantum,liu2008quantum,qian2014quantum,wang2014quantum,konig2007quantum}. In these systems, an external electric field $\vec{E}$ induces nonequilibrium coherence—i.e., electric-field-driven band mixing~\cite{culcer2017interband,sekine2017quantum,atencia2022semiclassical}—giving rise to a Hall conductivity described by the Kubo formalism~\cite{go2020orbital,sinitsyn2006charge,jungwirth2003dc}:
\begin{align} \label{fvdfva}
    \sigma_{H}&=-\frac{e^2\hbar}{\Omega} \sum_{\vec{k}\eta\neq\eta'}(f_{\vec{k}\eta}-f_{\vec{k}\eta'})\\
    &\times\text{Im}\left\{\frac{\langle \vec{k}\eta\vert \hat{v}^x_{\vec{k}}\vert \vec{k}\eta'\rangle \langle \vec{k}\eta\vert  \hat{v}^y_{\vec{k}}\vert \vec{k}\eta'\rangle}{(\epsilon_{\vec{k}\eta'}-\epsilon_{\vec{k}\eta}+i\Gamma)^2}\right\}. \notag
\end{align}
Here $e$ is the elementary charge, $\hbar$ the reduced Planck constant, $\Omega$ the system volume, $\hat{v}^i_{\vec{k}}$ the velocity operator, and $f_{\vec{k}\eta}=1/(e^{\beta\epsilon_{\vec{k}\eta}}+1)$ the Fermi–Dirac distribution with inverse temperature $\beta=1/(k_BT)$ and energy spectrum $\epsilon_{\vec{k}\eta}$.

\begin{figure}[t]
\begin{center}
\includegraphics[width=0.48\textwidth]{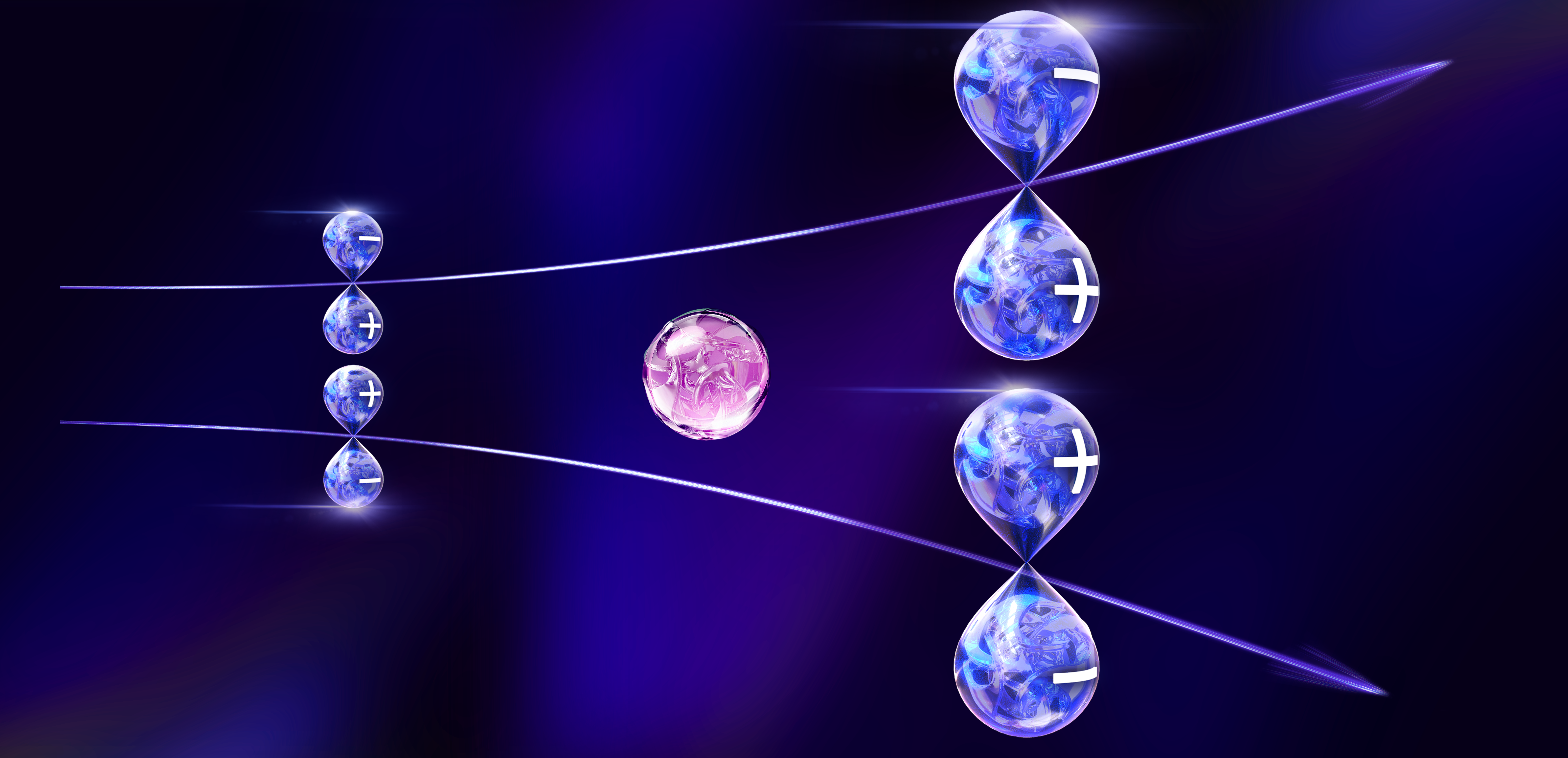} 
\end{center}
\caption{A second-order scattering process intrinsically tied to quantum decoherence. Electrons with ($-+$)  coherence scatter upward, while those with  ($+-$)  coherence tend to scatter downward.}
\label{FIG1}
\end{figure}

Nevertheless, quantum coherence is intrinsically fragile and progressively lost through interactions with the uncontrollable environment—such as magnetic impurities, hyperfine coupling, and electron–phonon or electron–electron scattering~\cite{chalker2007decoherence,jo2022scaling,nigg2016decoherence,jiang2009topological,qi2019dephasing}. This loss of coherence, i.e., decoherence, arises from phase randomization and energy dissipation processes~\footnote{Decoherence can be decomposed into two distinct pathways~\cite{zhang2024microscopic}: pure dephasing, which randomizes relative phases without altering populations, $\left\langle e^{ \text{ang}[c_{\alpha}(t)/c_{\alpha'}(t)]i} \right\rangle \sim e^{-t/\tau_\phi}$ and energy relaxation, which damps the magnitude of the superposition state~\eqref{fvdfvaml}, $\langle |c_{\alpha}(t)c^{*}_{\alpha}(t)| \rangle \sim e^{-t/\tau_1}$. The coherence magnitude then decays as $\langle |c_{\alpha}(t)c^{*}_{\alpha'}(t)| \rangle \sim e^{-t/(2\tau_1)}$, yielding the total decoherence time $\tau_2$ through $\frac{1}{\tau_2}=\frac{1}{2\tau_1}+\frac{1}{\tau_\phi}$, a standard result in quantum information and condensed-matter theory~\cite{breuer2002theory,nielsen2010quantum,abragam1961principles}.}, and drives the transition from quantum to classical behavior~\cite{breuer2002theory,nielsen2010quantum,zhang2024theory,zhang2024microscopic,zhang2025open,abragam1961principles}. Despite its central role, a transparent microscopic description of decoherence in spin–orbit–coupled ferromagnets remains lacking. Existing treatments are largely based on non-equilibrium Green’s-function approaches~\cite{nagaosa2010anomalous,sinitsyn2006charge,dugaev2005anomalous,crepieux2001theory,inoue2004suppression}, which, while formally rigorous, often obscure the underlying physical mechanisms.  In practice, decoherence is introduced phenomenologically into transport theory via a finite rate $\Gamma$ inserted into the Kubo formula [Eq.~\eqref{fvdfva}] and/or Green functions~\cite{zhou2022transport,jiang2009topological,go2020orbital,czaja2014anomalous,wan2025quantum}, with $\Gamma$ commonly treated as an adjustable parameter for fitting experiments~\cite{jungwirth2003dc}.

In this Letter, we develop a  master-equation framework with a general ansatz for the off-diagonal density matrix that captures both electric–field–induced coherence generation and impurity–scattering–driven coherence dissipation. This unified approach provides a quantitative description of how decoherence governs quantum transport in spin–orbit–coupled ferromagnets—an essential aspect for designing and operating functional quantum materials and devices. As a concrete application, we show that decoherence fundamentally modifies the AHE and uncover a distinct, ubiquitous contribution to the extrinsic Hall response: a second-order scattering mechanism originating from decoherence, different from both skew scattering and side jump mechanisms and often much stronger than the skew scattering.

\textit{Model and theory--}We study the nonequilibrium quantum kinetic equation of electrons, whose dynamics depends on the total Hamiltonian $\hat{H}=\hat{H}_{e}+\hat{H}_{E}+\hat{V}$ with $\hat{H}_{E}=-e\vec{E}\cdot \vec{r}$, and $\hat{V}=\sum_{j}U\delta(\vec{r}-\vec{R}_j)$, where $e<0$ is the charge of electron and $U$ denotes the scattering strength of the impurities located at random positions $\vec{R}_j$.  The  unperturbed Hamiltonian of electrons is $\hat{H}_{e}=\epsilon_{k}+v_{R} (k^xs^y-  k^ys^x)+\epsilon_L\hat{s}^z$ with $\epsilon_{k}=\hbar^2k^2/2m$,  where $\vec{k}=-i\vec{\nabla}_{\vec{r}}$ is canonical kinetic momentum of the electron at position $\vec{r}$, and $v_R$ quantifies the Rashba-type spin-orbit coupling (SOC).  The Zeeman term $\epsilon_L$ contains  the external Zeeman magnetic field and the spin-exchange field (SEF), and $\hat{\vec{s}}=(\hat{s}^x,\hat{s}^y,\hat{s}^z)$ is the vector of Pauli matrices of the  electron. Diagonalizing the unperturbed Hamiltonian generates the two isotropic energy bands $\epsilon_{k\eta}=\epsilon_{k}+\eta\mathcal{E}_{k}$, 
with $\mathcal{E}_{k}=\sqrt{ v^2_{R}k^2+\epsilon_{L}^2}$.  The associated eigenstates are  $\vert \vec{k}+ \rangle=\begin{bmatrix}
        \cos\frac{\Theta_{k}}{2}e^{-i\theta_{\vec{k}}}&
        +i\sin\frac{\Theta_{k}}{2}
    \end{bmatrix}^T$ and $\vert \vec{k}- \rangle=\begin{bmatrix}
        \sin\frac{\Theta_{k}}{2}e^{-i\theta_{\vec{k}}} &
        -i\cos\frac{\Theta_{k}}{2}
    \end{bmatrix}^T$, 
with $\cos\Theta_{k}=\epsilon_L/\mathcal{E}_{k}$, $\sin\Theta_{k}=v_{R}k/\mathcal{E}_{k}$, and $\theta_{\vec{k}}=\text{angle}\left(\frac{k_x+ik_y}{k}\right)$. Bellow, we show the external electric field ($\hat{H}_{E}$) and scalar scattering potential ($\hat{V}$), whose Hamiltonian do not commute with $\hat{H}_{e}$, induce band mixing, responsible for coherence generation [Eq.~\eqref{fvfvkdfk}] and dissipation [Eq.~\eqref{fdfkvIavf}], respectively.

\begin{figure}[t]
\begin{center}
\includegraphics[width=0.48\textwidth]{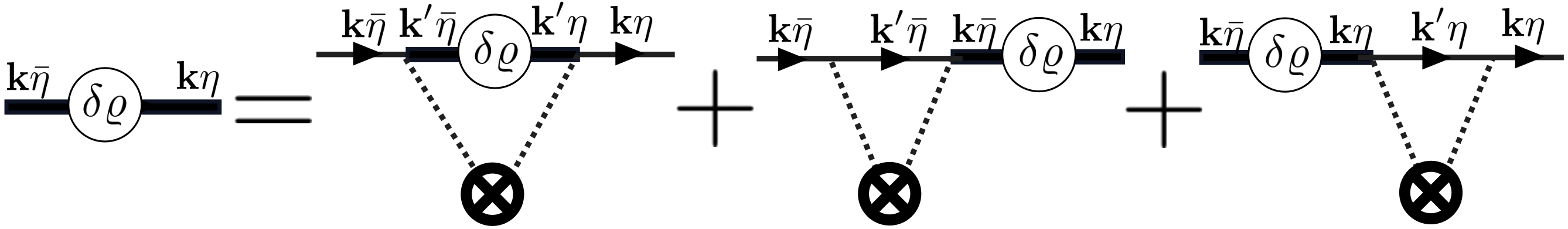} 
\end{center}
\caption{The Feynman diagrams for second-order Born-Markov approximation.}
\label{FIGD}
\end{figure}

For a weak external electric field $\vec{E}$, the linearized quantum kinetic equation, i.e., the time evolution of nonequilibrium off-diagonal density matrix~\cite{culcer2017interband,sekine2017quantum,atencia2022semiclassical}, is  
\begin{align} \label{fvdvkfvkmain1}
    \frac{\partial}{\partial t}  \delta\varrho^{\bar{\eta}\eta}_{\vec{k}}-\eta\frac{2i\mathcal{E}_{k}}{\hbar} \delta\varrho^{\bar{\eta}\eta}_{\vec{k}}+\mathcal{D}^{\bar{\eta}\eta}_{\vec{k}}(f)= \mathcal{ J}^{\bar{\eta}\eta}_{\vec{k}}(\delta\varrho),
\end{align}
with $\bar{\eta}=-\eta$. The external electric field causes energy bands mixes, i.e., electric field-induced coherence~\cite{culcer2017interband,sekine2017quantum,atencia2022semiclassical} 
\begin{align} \label{fvfvkdfk}
    \mathcal{D}^{\bar{\eta}\eta}_{\vec{k}}(f)=\frac{i}{\hbar}\vec{\mathcal{R}}^{\bar{\eta}\eta}_{\vec{k}} (f_{k\bar{\eta}}- f_{k\eta})\cdot e\vec{E},
\end{align}
where only off-diagonal components of the Berry connection participate in decoherence generation in the drive term~\eqref{fvfvkdfk}, and are expanded in the eigen basis  $\mathcal{R}^{\bar{\eta}\eta}_{x,\vec{k}}=\frac{\sin\Theta_{k}}{2k}\left[-\sin \theta_{\vec{k}}-\cos\Theta_{k}\cos\theta_{\vec{k}}\eta^y_{\vec{k}}\right]$ and  $\mathcal{R}^{\bar{\eta}\eta}_{y,\vec{k}}=\frac{\sin\Theta_{k}}{2k}\left[+\cos \theta_{\vec{k}}\eta^x_{\vec{k}}-\cos\Theta_{k}\sin\theta_{\vec{k}}\eta^y_{\vec{k}}\right]$. In the presence of the randomly distributed impurities, the interband collision integral, within second-order \textit{Born-Markov approximation}, is given by \cite{breuer2002theory,zhang2024theory,zhang2024microscopic,zhang2025open}
\begin{align} \label{fdfkvIavf}
   \mathcal{ J}^{\bar{\eta}\eta}_{\vec{k}}(\delta\varrho)&\simeq\frac{\pi n_{\text{i}}U^{2}}{\hbar \Omega}\sum_{\vec{k}'} \left\lbrace 
   \delta(\epsilon_{k'\eta}-\epsilon_{k\eta}) V_{\vec{k}\bar{\eta},\vec{k}'\bar{\eta}} \delta\varrho_{\vec{k}'}^{\bar{\eta}\eta} V_{\vec{k}'\eta,\vec{k}\eta}      
   \right. \notag\\
   &+\delta(\epsilon_{k\bar{\eta}}-\epsilon_{k'\bar{\eta}})  V_{\vec{k}\bar{\eta},\vec{k}'\bar{\eta}}   \delta\varrho_{\vec{k}'}^{\bar{\eta}\eta} V_{\vec{k}'\eta,\vec{k}\eta}\notag\\
   &-\delta(\epsilon_{k'\bar{\eta}}-\epsilon_{k\bar{\eta}}) V_{\vec{k}\bar{\eta},\vec{k}'\bar{\eta}}  V_{\vec{k}'\bar{\eta},\vec{k}\bar{\eta}}  \delta\varrho_{\vec{k}}^{\bar{\eta}\eta}\notag\\
&-\left. \delta(\epsilon_{k\eta}-\epsilon_{k'\eta}) \delta\varrho_{\vec{k}}^{\bar{\eta}\eta} V_{\vec{k}\eta,\vec{k}'\eta}   V_{\vec{k}'\eta,\vec{k}\eta}%
   \right\}.
\end{align}
Both $ \delta(\epsilon_{k'\eta}-\epsilon_{k\eta})$ and $V_{\vec{k}'\eta',\vec{k}\eta}=\langle \vec{k}'\eta'\vert s^o\vert \vec{k}\eta \rangle $ incorporate the effects of the SOC and SEF on impurity scattering. Here, we have excluded interband scattering processes which are energetically forbidden in the weak-disorder limit and the Feynman diagrams for second-order Born-Markov approximation is plotted in Fig.~\ref{FIGD}. 
The first and second terms (positive sign) of the collision integral \eqref{fdfkvIavf}, plotted by the first Feynman diagram, describe scattering-in processes (gain term), where an electron transitions from a $\vec{k}'$ state  to the $ \vec{k}$ state, thereby increasing the coherence $\delta\varrho^{\bar{\eta}\eta}_{\vec{k}}$. Conversely, the third and fourth ones  (negative sign), plotted by the second and third Feynman diagrams, describe scattering-out processes  (loss term), where an electron leaves the $\vec{k}$ state for another $\vec{k}'$ state, leading to a decrease in the  coherence $\delta\varrho^{\bar{\eta}\eta}_{\vec{k}}$. The  collision integral~\eqref{fdfkvIavf}, which vanishes at thermal equilibrium, scales with the deviation from thermal equilibrium, and at steady state, the field-induced coherence generation~\eqref{fvfvkdfk} is exactly balanced by collisional coherence dissipation~\eqref{fdfkvIavf}.

\textit{AHE from coherence--}Unlike classical drift–diffusion formula, quantum transport must account for quantum coherence over the device length. In a clean enough material, impurity scattering is ignorable, i.e., $\mathcal{ J}^{\eta_2\eta_1}_{\vec{k}}(\delta\varrho)=0$. Then, the nonequilibrium off-diagonal  density matrix, in steady state ($\partial_t\delta\varrho^{\bar{\eta}\eta}_{\vec{k}}$=0), is  proportional to the Berry connection ($\vec{\mathcal{R}}^{\bar{\eta}\eta}_{\vec{k}}$) and can be derived from Eq.~\eqref{fvdvkfvkmain1} 
\begin{align} \label{agflalgt}
    \delta\varrho^{\bar{\eta}\eta}_{\vec{k}}=-\frac{ f_{k\bar{\eta}}- f_{k\eta}}{\epsilon_{k\bar{\eta}}- \epsilon_{k\eta}}\vec{\mathcal{R}}^{\bar{\eta}\eta}_{\vec{k}}\cdot e\vec{E},
\end{align}
which encodes perfect coherence, (i.e., $\tau_{\phi}=\tau_{1}=\infty$).  
The quantum coherence, quantified by $\delta\varrho^{\bar{\eta}\eta}_{\vec{k}}$, contributes to a charge current in transverse direction, thus accounting for intrinsic AHE~\cite{culcer2017interband}. 
The associated charge Hall conductivity, i.e.,  the Kubo formula \eqref{fvdfva} with $\Gamma=0$, is  
\begin{align} \label{bgbgbsbg}
    \sigma^{0}_{H}= \frac{e^2}{2h} \sum_{\eta }\eta\frac{ \epsilon_L}{ \sqrt{(v_Rk^{\eta}_F)^2+\epsilon_L^2}} ,
\end{align}
where $k^{\eta}_F$, relying on the Fermi energy, is the Fermi kinetic momentum for $\eta$ band, i.e., $\epsilon_{k^{\eta}_F\eta}=\epsilon_F$, with $\epsilon_F$ being the Fermi energy.  Hereafter, we assume $\epsilon_F > -\vert \epsilon_L \vert$~\footnote{Distinct behaviors emerge depending on whether the Fermi energy $\epsilon_F$ lies above or below $\vert \epsilon_L \vert$.}. However, experimental measurements typically deviate from the perfect coherence result, i.e., Eq.~\eqref{bgbgbsbg}, due to the unavoidable presence of decoherence.  


\textit{Theory of decoherence--}In a dirty material, impurity-induced electron scattering becomes appreciable, and thus we are required to include the collision integral~\eqref{fdfkvIavf}.  There exist two extrinsic mechanisms of AHE, stemming from skew scattering~\cite{sinitsyn2007anomalous,schliemann2003anisotropic} and side jump~\cite{berger1970side,sinitsyn2006coordinate}, respectively. Both are calculated from the quantum distribution--the diagonal component of the density matrix. Here, we work on a new extrinsic mechanism from the decoherence -- the relaxation of the off-diagonal density matrix, which goes beyond the previous results of the perfect coherence~\cite{culcer2017interband}. We use an ansatz of the off-diagonal density matrix 
\begin{align} \label{ansatz}
    \delta\varrho^{\bar{\eta}\eta}_{\vec{k}}=\delta\varrho^{\bar{\eta}\eta}_{\vec{k},\Vert}+\delta\varrho^{\bar{\eta}\eta}_{\vec{k},\perp},
\end{align}
with
\begin{align} \label{fvfkvmkdf4}
\delta\varrho^{\bar{\eta}\eta}_{\vec{k},\Vert/\perp}=-\frac{e}{\hbar}\tau^{\bar{\eta}\eta}_{k,\Vert/\perp}(f_{k\bar{\eta}}- f_{k\eta})\vec{\mathcal{R}}^{\bar{\eta}\eta}_{\vec{k}}\cdot \vec{E}_{\Vert/\perp},
\end{align}
where $\vec{E}_{\Vert}=\vec{E}$ and $\vec{E}_{\perp}=\vec{E}\times \hat{z}$. Our ansatz \eqref{ansatz} recovers  the purely intrinsic case  \eqref{agflalgt} with  $\tau^{\bar{\eta}\eta}_{k,\Vert}=\hbar/(\epsilon_{k\bar{\eta}}- \epsilon_{k\eta})$ and $\tau^{\bar{\eta}\eta}_{k,\perp}=0$. Importantly, a proper ansatz of the off-diagonal density matrix is crucial for derivations of collision integral \eqref{fdfkvIavf}, from that we extract decoherence rate (see details in Appendix A).  Here, the $\delta\varrho^{\bar{\eta}\eta}_{\vec{k},\Vert}$ component is inspired by the drive term~\eqref{fvfvkdfk}, and we then show $\delta\varrho^{\bar{\eta}\eta}_{\vec{k},\perp}$ component is required by the \textit{second-order} scattering processes tightly relative to quantum coherence.

Notably, to include decoherence, our collision term must go beyond the Fermi's Golden rule within diagonal approximations of density matrix and collision integral, i.e., $\delta\varrho_{\vec{k}}^{\eta_2\eta_1}=\delta_{\eta_2\eta_1}\delta\varrho_{\vec{k}}^{\eta_1\eta_1}$ and $\mathcal{ J}^{\eta_2\eta_1}_{\vec{k}}(\delta\varrho)=\delta_{\eta_2\eta_1}\mathcal{ J}^{\eta_1\eta_1}_{\vec{k}}(\delta\varrho)$, where the scattering rate $\propto \vert V_{\vec{k}\eta,\vec{k}'\eta'}\vert^2$ is real and symmetric in the incident and scattered electrons: the magnitude for scattering from $\vert \vec{k} \eta\rangle $ to $\vert \vec{k}' \eta'\rangle $ is the same as the magnitude for
scattering from $\vert \vec{k}' \eta'\rangle$ to $\vert \vec{k} \eta\rangle$. Importantly, the collision integral \eqref{fdfkvIavf} incorporates the off-diagonal density matrix $\delta\varrho_{\vec{k}'}^{\bar{\eta}\eta}$, and thus there exist several \textit{second-order} scattering-in processes associated with quantum coherence whose scattering rate $V_{\vec{k}\bar{\eta},\vec{k}'\bar{\eta}}  V_{\vec{k}'\eta,\vec{k}\eta}$ has an antisymmetric component, i.e.,  $\sin\theta_{\vec{k}'\vec{k}}$ [3rd line of Eq.~\eqref{gbafkgbfg} in Appendix A], unlike typical second-order scattering rate $\vert V_{\vec{k}\eta,\vec{k}'\eta'}  \vert^2$ within the Fermi golden rule. 
This type of second-order scattering relative to quantum coherence is fundamentally distinct from, yet substantially more significant than, the third-order skew scattering associated with the diagonal density matrix~\cite{Ma2023anomalous,culcer2022anomalous}. When evaluating the collision integral \eqref{fdfkvIavf}, the integration of the gain terms $V_{\vec{k}\bar{\eta},\vec{k}'\bar{\eta}}  V_{\vec{k}'\eta,\vec{k}\eta} \delta\varrho^{\bar{\eta}\eta}_{\vec{k}'}$ over $\theta_{\vec{k}'}$ couples the ordinary off-diagonal density matrix $\delta\varrho^{\bar{\eta}\eta}_{\vec{k}',\Vert}$ to its anomalous counterpart $\delta\varrho^{\bar{\eta}\eta}_{\vec{k},\perp}$ due to the identity
\begin{align} \label{trfvdvk1}
    \int^{2\pi}_{0}\sin\theta_{\vec{k}'\vec{k}}\vec{\mathcal{R}}_{\vec{k}'\eta_i}\cdot \vec{E}_{\Vert}=\pi \vec{\mathcal{R}}_{\vec{k}\eta_i}\cdot \vec{E}_{\perp}.
\end{align}
This explains why we add $\delta\varrho^{\bar{\eta}\eta}_{\vec{k},\perp}$ component in our anstaz~\eqref{ansatz}, serving as a transverse drive field ($\vec{E}_{\perp}=\vec{E}\times \hat{z}$). Thus, our \textit{second-order} scattering  tightly relative to decoherence acts as an effective out-of-plane magnetic field, exerting a Lorentz force perpendicular to the direction of current flow. 

Following the detailed derivations in Appendix A, we attain the interband collision term as follows: 
\begin{align} \label{fvafvf}
     \mathcal{ J}^{\bar{\eta}\eta}_{\vec{k}}=-\frac{1}{\hbar}\Gamma_{k} \delta\varrho^{\bar{\eta}\eta}_{\vec{k}}+i\eta\frac{1}{\hbar}\Gamma^a_{k}\delta\varrho^{\bar{\eta}\eta}_{\vec{k},a},
\end{align}
with
\begin{align} \label{gbbkgbm}
    \delta\varrho^{\bar{\eta}\eta}_{\vec{k},a}=\frac{\tau^{\bar{\eta}\eta}_{k,\Vert}}{\tau^{\bar{\eta}\eta}_{k,\perp}}\delta\varrho^{\bar{\eta}\eta}_{\vec{k},\perp}-\frac{\tau^{\bar{\eta}\eta}_{k,\perp}}{\tau^{\bar{\eta}\eta}_{k,\Vert}}\delta\varrho^{\bar{\eta}\eta}_{\vec{k},\Vert}.
\end{align}
The imaginary, antisymmetric components carry opposite signs for different values of $\eta$. Since $\delta\varrho_{\bar{\eta}\eta}$ quantifies the electron coherence—the superposition of  $|\vec{k},\bar{\eta}\rangle$  and  $|\vec{k},\eta\rangle$—the scattering probabilities to the top and bottom directions become unequal. As illustrated in Fig.~\ref{FIG1}, electrons with ($-+$)  coherence preferentially scatter upward, whereas those with  ($+-$) coherence favouredly tend to scatter downward. Using the collision integral~\eqref{fvafvf}, the quantum kinetic equation~\eqref{fvdvkfvkmain1} is separated into ordinary and anomalous components of density matrix, i.e., $\delta\varrho^{\bar{\eta}\eta}_{\vec{k},\Vert}$ and $\delta\varrho^{\bar{\eta}\eta}_{\vec{k},\perp}$. Solving them, in steady state ($ \partial_t  \delta\varrho^{\bar{\eta}\eta}_{\vec{k}}$=0), gives rise to the normal and anomalous decoherence time 
\begin{align} \label{sgfbf1}
    \frac{\tau^{\bar{\eta}\eta}_{k,\Vert}}{\hbar}=\frac{(\epsilon_{k\bar{\eta}}-\epsilon_{k\eta})-i\Gamma_{k}}{[(\epsilon_{k\bar{\eta}}-\epsilon_{k\eta})-i\Gamma_{k}]^2+(\Gamma^a_{k})^2},
\end{align}
\begin{align} \label{sgfbf2}
    \frac{\tau^{\bar{\eta}\eta}_{k,\perp}}{\hbar}=\frac{\eta\Gamma^a_{k}}{[(\epsilon_{k\bar{\eta}}-\epsilon_{k\eta})-i\Gamma_{k}]^2+(\Gamma^a_{k})^2}.
\end{align}
The ordinary scattering causes decoherence, i.e., the decay of the off-diagonal density matrix, which is quantified by a normal decoherence rate $\Gamma_{k}/\hbar=\frac{1}{8}\left(1+\frac{\epsilon^2_L}{\mathcal{E}^2_k}\right)\left(\frac{1}{\tau^{0}_{k+}}+\frac{1}{\tau^{0}_{k-}}\right)$, where $1/\tau^0_{k\eta}=(2\pi/\hbar)n_{\text{i}}\nu_{\eta}(\epsilon_{k\eta})U^{2}$, which reduces to $1/\tau^0=\frac{m n_{\text{i}}U^{2}}{\hbar^3}$ in the absence of the SOC. 
The density of state for each energy band is given by $\nu_{\eta}(\epsilon)=\left.\frac{1}{h}\frac{\kappa}{\left\vert v_{\kappa\eta}\right\vert}\right\vert_{\epsilon=\epsilon_{\kappa\eta}}$  with $v_{k\eta}=\frac{\hbar k}{m}+\eta\frac{v_R}{\hbar}\sin\Theta_{k}$. More importantly, the anomalous scattering, quantified by an anomalous decoherence rate $\Gamma^a_{k}/\hbar= \frac{\epsilon_L}{4\mathcal{E}_k}  \left(\frac{1}{\tau^{0}_{k+}}+\frac{1}{\tau^{0}_{k-}}\right) $, generates an effective out-of-plane magnetic field [see Fig.~\ref{FIG1}]. 
The decoherence rate $\Gamma_{k}$ and $\Gamma^a_{k}$, which are of order $ \hbar/\tau^0$~\footnote{In the absence of SOC, we have $\Gamma_{k}=\hbar/(2\tau^0)$ and $\Gamma^a_{k}=\hbar/(2\tau^0)$.}, should play an important role in the quantum transport associated quantum coherence (i.e., off-diagonal density matrix), like the momentum relaxation -- the lifetime of quantum population (i.e., diagonal density matrix). We have demonstrated a decoherence-induced conductivity \textit{linear} in $n_{\text{i}}$, markedly contrasting the conventional Drude conductivity inversely proportional to $n_{\text{i}}$ and indicating that small amount of impurities can produce observable  magnetoresistance from decoherence~\cite{zhang2026magnetoresistance}. Here, we study the effect of decoherence on the AHE.

\begin{figure}[t]
\begin{center}
\includegraphics[width=0.48\textwidth]{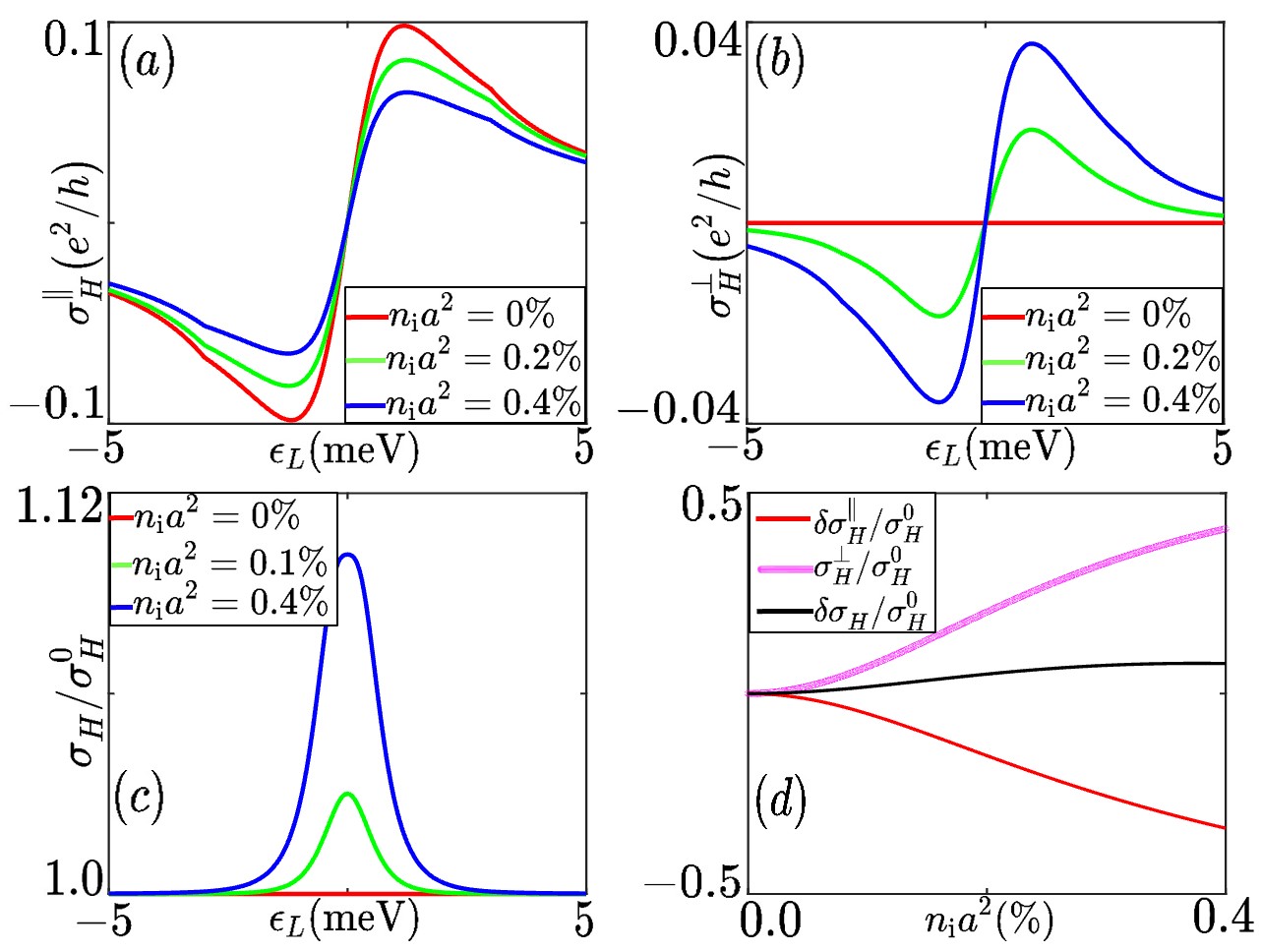}   
\end{center}
\caption{(a-c) The $\epsilon_L$ dependence of the AHE conductivities derived from (a) ordinary, (b) anomalous, and (c) total off-diagonal density matrix. Panel (d) plots the $n_{\text{i}}$ dependence of $\delta\sigma_{H}$, $\delta \sigma^{\Vert}_{H}$, and  $\sigma^{\perp}_{H}$. Other parameters: $v_R/a=12$ meV, and  $\epsilon_F=3$ meV, $U/a^2=1$ eV, and $a=0.35$ nm.}
\label{FIG}
\end{figure}

\textit{Influence of decoherence on AHE} --
Let us begin with the contribution of $\delta\varrho^{\bar{\eta}\eta}_{\vec{k},\Vert}$.  The real part of $\delta\varrho^{\bar{\eta}\eta}_{\vec{k},\Vert}$ contributes to a charge Hall  conductivity
\begin{align} \label{oHov}
    \sigma^{\Vert}_{H}&=-\frac{e^2}{ h} \frac{v_R }{\hbar} \int^{k^{+}_F}_{k^{-}_F} dk \frac{v_Rk}{\mathcal{E}_k}\frac{\epsilon_L}{\mathcal{E}_k}\text{Re} (\tau^{+-}_{k,\Vert}),
\end{align}
which reduces to Eq.~\eqref{bgbgbsbg} at $n_{\text{i}}=0$. In the dilute limit, the real part of the normal decoherence time \eqref{sgfbf1} reduces to $\text{Re}(\tau^{+-}_{k,\Vert}/\hbar)\simeq 1/(2\mathcal{E}_k)-[\Gamma^2_k+(\Gamma^a_k)^2]/(8\mathcal{E}^3_k)$. Thus, the decoherence-induced correction of $\sigma^{\Vert}_{H}$, quantified by the difference $\delta \sigma^{\Vert}_{H}=\sigma^{\Vert}_{H}-\sigma^{0}_{H}$, grows quadratically with $n_{\text{i}}$ [see also Fig.~\ref{FIG}(d)], indicating that appreciable impurity density is required to produce observable decoherence effects. Quantitatively, Figure \ref{FIG}(a) plots $\sigma_{H}^{\Vert}$  for different values of $n_{i}$. The SEF enhances decoherence by increasing both the normal and anomalous decoherence rates, i.e., $\Gamma^a_k (\propto \epsilon_L)$ and $\Gamma_k \propto \left(1 + \epsilon_L^2 / \mathcal{E}_k^2\right)$. As a result, decoherence suppresses the Hall conductivity more efficiently at larger $\epsilon_L$. However, in the strong-SEF regime ($\epsilon_L \gg v_R/a$), coherent transport itself is quenched, and decoherence effects consequently vanish. These results highlight the crucial role of decoherence in the AHE.

\textit{Decoherence mechanism of AHE--}Coherent quantum transport conventionally requires minimizing decoherence. Here, we act in a diametrically opposite and demonstrate that decoherence itself can give rise to the AHE. Specially,  $\delta\varrho^{\bar{\eta}\eta}_{\vec{k},\perp}$ also contributes to a  charge Hall conductivity reads 
\begin{align} \label{oHop}
    \sigma^{\perp}_{H}=+\frac{e^2}{ h} \frac{v_R}{2\hbar} \int^{k^{+}_F}_{k^{-}_F} dk\left(1+\frac{\epsilon^2_L}{\mathcal{E}^2_k}\right)\frac{v_Rk}{\mathcal{E}_k}\text{Im} (\tau^{+-}_{k,\perp}).
\end{align}
The imaginary part of the anomalous decoherence time \eqref{sgfbf2}, in the dilute limit, reduces to $\text{Im}(\tau^{+-}_{k,\perp}/\hbar)\simeq -\Gamma_k\Gamma^a_k/(4\mathcal{E}^3_k)$. Again, the AHE governed by $\tau^{+-}_{k,\perp}$ grows quadratically with $n_{\text{i}}$ (i.e., $ \sigma^{\perp}_{H} \propto n^2_{\text{i}}$). This decoherence-driven AHE vanishes entirely in the absence of decoherence [see red curves in Fig. \ref{FIG}(b)], as the anomalous decoherence rate ($\Gamma^a_k \propto n_{\text{i}}$) approaches zero. Increasing $n_{\text{i}}$, causes sizable value of $\Gamma^a_k$ and generates a measurable AHE conductivity, $\sigma^\perp_{H}/\sigma^0_{H} \sim 0.41,$ at $\epsilon_L=0.5$ meV [the blue curve of Fig. \ref{FIG}(b)].  This reveals a distinct and ubiquitous  mechanism for the extrinsic AHE that originates from the \textit{second-order} scattering processes owing to decoherence—that is fundamentally distinct from two extrinsic mechanisms--skew scattering and side jump~\cite{culcer2022anomalous,nagaosa2010anomalous,sinitsyn2007anomalous}.

Notably, $\delta\sigma^{\Vert}_{H}$ and $\sigma^{\perp}_{H}$ have opposite signs, leading to a partial cancellation of the two decoherence-induced corrections, as shown by the red and magenta curves of Fig.~\ref{FIG}(d). As a result, the total correction to the AHE conductivity, $\delta \sigma_{H}=\delta\sigma^{\Vert}_H+\sigma^{\perp}_H$, remains robust against variations in the impurity density [Figs.~\ref{FIG}(c) and (d)]. Importantly, decoherence in fact, enhances the total AHE effect, i.e., $\sigma_H/\sigma^0_H>1$, as demonstrated in Figs.~\ref{FIG}(c).  In contrast, the conventional Kubo formula \eqref{fvdfva}, which introduces a phenomenological decoherence rate $\Gamma$, suffers from two limitations: (i) it fails to capture the $k$-dependence of the decoherence time, and (ii) it overestimates the effect of decoherence by neglecting $\sigma^{\perp}_H$ (e.g., $\sigma^\perp_{H}/\sigma^{0}_{H} \sim 0.41,$ at $\epsilon_L=0.5$ meV). Consequently, our formulism is crucial for a quantitative analysis of decoherence effects on the AHE.

\begin{figure}
\begin{center}
\includegraphics[width=0.48\textwidth]{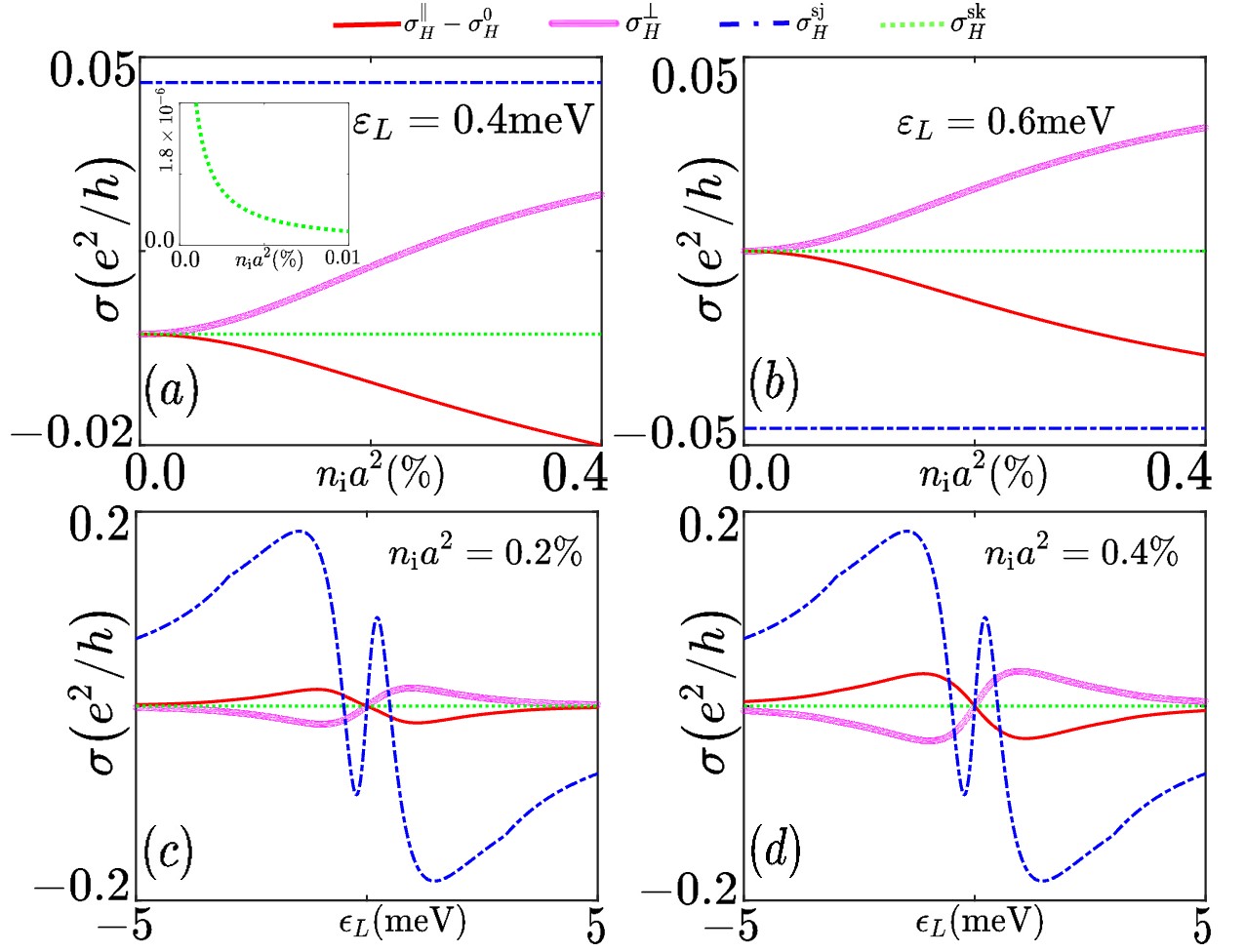} 
\end{center}
\caption{(a-b) $n_{\text{i}}$ dependence of the AHE conductivities derived from side jump (blue), skew scattering (green), and anomalous (magenta) off-diagonal density matrix. The red curves reveal a clear influence of decoherence on the AHE, quantified by $\sigma^{\Vert}_{H}-\sigma^{0}_{H}$. Panels (c) and (d) plot the corresponding $\epsilon_L$ dependence.  The inset of panel (a) reveals the $1/n_{\textit{i}}$ scaling of the skew scattering contribution.
Other parameters are the same as Fig. \ref{FIG}.}
\label{FIG1N}
\end{figure}

\begin{figure}[b!]
\begin{center}
\includegraphics[width=0.48\textwidth]{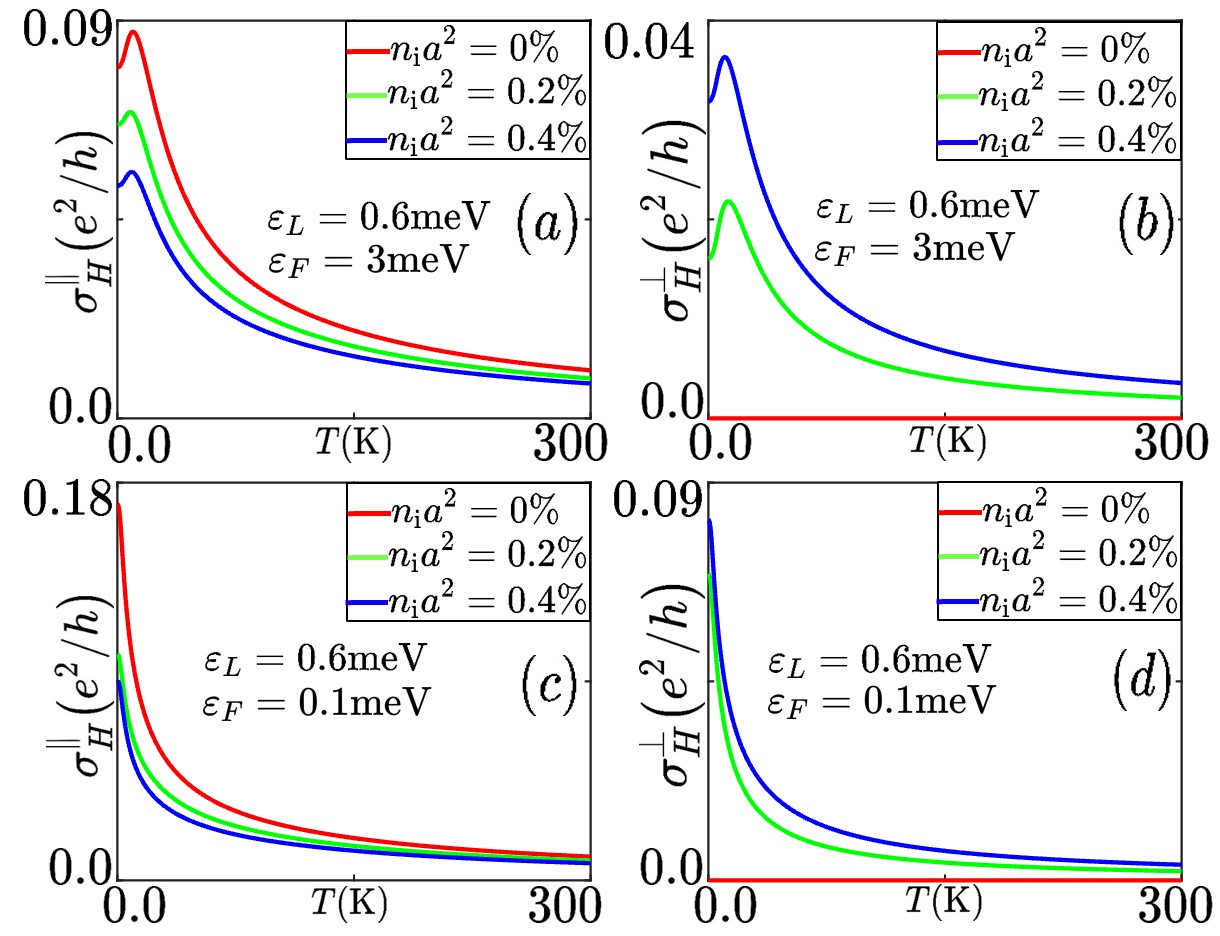} 
\end{center}
\caption{(a,b) $T$ dependence of (a) $\sigma^{\Vert}_{H}$ and (b) $\sigma^{\perp}_{H}$ for different   $n_{\text{i}}$, where $\epsilon_F>\epsilon_L$. Panels (c) and (d) plot the situation of $\epsilon_F<\epsilon_L$.  Other parameters are the same as Fig. \ref{FIG}.}
\label{FIG2}
\end{figure}

\textit{Scaling analysis--} Since the transport lifetime scales as $\tau\propto 1/n_{\mathrm{i}}$ and the Drude conductivity follows $\sigma_{xx}\propto1/n_{\mathrm{i}}$, the decoherence-induced AHE contributions ($\delta\sigma_H^{\parallel}$, $\sigma_H^{\perp}$, or $\delta\sigma_H$) scale quadratically with $n_{\mathrm{i}}$ in the dilute limit [Fig.~\ref{FIG}(d)]. This quadratic scaling agrees with the early Green’s-function results of Alexei Abrikosov~\cite{abrikosov1969galvanomagnetic}, originally developed for the ordinary Hall effect in strong orbital magnetic fields. 
Experimentally, the unconventional scaling should be an inevitable outcome in Berry-curvature-dominated systems, and has indeed been reported in ferromagnets~\cite{peng2025unconventional,feng2020nonvolatile} and lanthanum doped NiCo$_2$O$_4$ film~\cite{zhang2024tunable}.  These impurity-dependent signatures provide a clear experimental diagnostic that distinguishes our mechanisms from intrinsic and side-jump contributions that are insensitive to $n_{\text{i}}$~\cite{sinova2015spin,chang2023colloquium}, as well as skew scattering contribution that scales with $1/n_{\text{i}}$ [inset of Fig.~\ref{FIG1N}(a)]. Furthermore, the resulting Hall conductivity is parametrically larger than the skew-scattering contribution (Fig.~\ref{FIG1N}), enabling substantially stronger and experimentally accessible modulation of the AHE with magnetic field and temperature (Figs.~\ref{FIG1N} and \ref{FIG2}). These features underscore the conceptual and practical distinctiveness of the decoherence-enabled extrinsic mechanism.

\textit{Conclusion--} In summary, we have developed a general ansatz for the off-diagonal density matrix that captures the generation and dissipation of quantum coherence under electric fields and impurity scattering, while fully incorporating the interplay between SOC and SEF. This framework reveals how decoherence modifies the intrinsic AHE and mediates a crossover between intrinsic and extrinsic regimes. Notably, we identify a new extrinsic contribution—a second-order scattering process enabled by decoherence—that is fundamentally different from both skew scattering and side-jump mechanisms, and significantly larger than the skew-scattering mechanism. Our approach provides a unified foundation for quantum transport in spin–orbit–coupled systems and can be readily extended to topological insulator~\cite{zhang2026resistance,zhang2026quantum} and materials with anisotropic spin splitting, such as altermagnets, where quantum decoherence also governs magnetoresistance and spin/valley/orbital Hall effects~\cite{zhang2026magnetoresistance,zhang2026quantum}.

\textit{Acknowledgement}--This work is supported by National Key R$\&$D Program of China (Grant Nos. 2020YFA0308800, 2021YFA1401500), the National Natural Science Foundation of China (Grant Nos. 12234003, W2511003, 12321004, 12022416, 12475015, 11875108), and the National Council for Scientific and Technological Development (Grant No. 301595/2022-4), and the Key Field Projects in Guangdong Province's Higher Educations Institution (Grant No. 2024ZDZX3066).

\newpage
\onecolumngrid
\vspace{1em}
\begin{center}
	\textbf{\large End Matter}
\end{center}
\vspace{1em}
\twocolumngrid

\textit{Appendix A: Derivations of decoherence--} 
We analytically derive the collision integral based on our ansatz~\eqref{ansatz}. Here, we include only intraband scattering processes, since direct interband scattering is energetically forbidden in the weak-disorder limit of our model~\cite{Datta1995,Bruus2004}.  The resulting interband collision integral, which merely incorporates the off-diagonal density matrix, is derived 
\begin{align} \label{rrefdfvrvolqplf1}
   \mathcal{ J}^{\bar{\eta}\eta}_{\vec{k}}(\delta\varrho)&=\frac{\pi n_{\text{i}}U^{2}}{\hbar \Omega}\sum_{\vec{k}'} \left\lbrace 
   +\delta(\epsilon_{k'\eta}-\epsilon_{k\eta}) V_{\vec{k}\bar{\eta},\vec{k}'\bar{\eta}} \delta\varrho_{\vec{k}'}^{\bar{\eta}\eta} V_{\vec{k}'\eta,\vec{k}\eta}      
   \right. \notag\\
   &-\delta(\epsilon_{k\eta}-\epsilon_{k'\eta}) \delta\varrho_{\vec{k}}^{\bar{\eta}\eta} V_{\vec{k}\eta,\vec{k}'\eta}   V_{\vec{k}'\eta,\vec{k}\eta}\notag\\
   &-\delta(\epsilon_{k'\bar{\eta}}-\epsilon_{k\bar{\eta}}) V_{\vec{k}\bar{\eta},\vec{k}'\bar{\eta}}  V_{\vec{k}'\bar{\eta},\vec{k}\bar{\eta}}  \delta\varrho_{\vec{k}}^{\bar{\eta}\eta}\notag\\
&+\left.\delta(\epsilon_{k\bar{\eta}}-\epsilon_{k'\bar{\eta}})  V_{\vec{k}\bar{\eta},\vec{k}'\bar{\eta}}   \delta\varrho_{\vec{k}'}^{\bar{\eta}\eta} V_{\vec{k}'\eta,\vec{k}\eta} %
   \right\}.
\end{align}
This is the collision integral, i.e., Eq.~\eqref{fdfkvIavf} in the main text.
The delta function in the collision integral above ensures energy conservation throughout the scattering processes. The first and second terms on the right-hand side of Eq.~\eqref{rrefdfvrvolqplf1} correspond to the scattering events associated with energy band $\eta$, while the third and fourth ones correspond to the energy band $\bar{\eta}$. Substituting ansatzes \eqref{fvfkvmkdf4} into Eq.~\eqref{rrefdfvrvolqplf1}, the collision integral can be further evaluated as follows: 
\begin{align} \label{gbafkgbfg}
     \mathcal{ J}^{\bar{\eta}\eta}_{\vec{k}}&=- \frac{\pi^2n_{\text{i}}U^{2}}{4\pi^2\hbar^2} \left(\frac{k}{ \left\vert v^{0}_{k+} \right\vert}+\frac{k}{ \left\vert v^{0}_{k-} \right\vert}\right) \left(1+\frac{\epsilon^2_L}{\mathcal{E}^2_{k}}\right) \delta\varrho_{\vec{k}}^{\bar{\eta}\eta} \notag \\
      &+ \frac{\pi n_{\text{i}}U^{2}}{8\pi^2\hbar^2} \left(\frac{k}{ \left\vert v^{0}_{k+} \right\vert}+\frac{k}{ \left\vert v^{0}_{k-} \right\vert}\right) \\
     &\times\int_{0}^{2\pi} d\theta_{\vec{k}'}\left[\left(1+\frac{\epsilon^2_L}{\mathcal{E}^2_{k}}\right)\cos\theta_{\vec{k}'\vec{k}}+2\eta \frac{\epsilon_L}{\mathcal{E}_{k}}i\sin\theta_{\vec{k}'\vec{k}}\right]\delta\varrho_{\vec{k}'}^{\bar{\eta}\eta}.\notag
\end{align}
Notably, the off-diagonal density matrix in the third line of Eq.~\eqref{gbafkgbfg} contains both $\delta\varrho^{\bar{\eta}\eta}_{\vec{k}',\Vert}$ and $\delta\varrho^{\bar{\eta}\eta}_{\vec{k}',\perp}$ components, which become $\delta\varrho^{\bar{\eta}\eta}_{\vec{k},\Vert}$ and $\delta\varrho^{\bar{\eta}\eta}_{\vec{k},\perp}$  after integrate with $\cos\theta_{\vec{k}'\vec{k}}$, while becomes $\delta\varrho^{\bar{\eta}\eta}_{\vec{k},\perp}$ and $\delta\varrho^{\bar{\eta}\eta}_{\vec{k},\Vert}$  after integrate with $\sin\theta_{\vec{k}'\vec{k}}$. We thus obtain:
\begin{align} \label{fdavfvd}
    \mathcal{ J}^{\bar{\eta}\eta}_{\vec{k}}&= - \frac{n_{\text{i}}U^{2}}{8\hbar^2} \left(\frac{k}{ \left\vert v^{0}_{k+} \right\vert}+\frac{k}{ \left\vert v^{0}_{k-} \right\vert}\right) \left(1+\frac{\epsilon^2_L}{\mathcal{E}^2_{k}}\right)\delta\varrho_{\vec{k}}^{\bar{\eta}\eta} \notag  \\
     &+ \eta i \frac{ n_{\text{i}}U^{2}}{4\hbar^2} \frac{\epsilon_L}{\mathcal{E}_{k}} \left(\frac{k}{ \left\vert v^{0}_{k+} \right\vert}+\frac{k}{ \left\vert v^{0}_{k-} \right\vert}\right) \delta\varrho_{\vec{k},a}^{\bar{\eta}\eta}, 
\end{align}
with
\begin{align} \label{gbbkgbm}
    \delta\varrho^{\bar{\eta}\eta}_{\vec{k},a}=\frac{\tau^{\bar{\eta}\eta}_{k,\Vert}}{\tau^{\bar{\eta}\eta}_{k,\perp}}\delta\varrho^{\bar{\eta}\eta}_{\vec{k},\perp}-\frac{\tau^{\bar{\eta}\eta}_{k,\perp}}{\tau^{\bar{\eta}\eta}_{k,\Vert}}\delta\varrho^{\bar{\eta}\eta}_{\vec{k},\Vert}.
\end{align}
Thus, we attain the interband collision term resulting from intraband scattering processes: 
\begin{align}
     \mathcal{ J}^{\bar{\eta}\eta}_{\vec{k}}=-\frac{1}{\hbar}\Gamma_{k} \delta\varrho^{\bar{\eta}\eta}_{\vec{k}}+i\eta\frac{1}{\hbar}\Gamma^a_{k}\delta\varrho^{\bar{\eta}\eta}_{\vec{k},a}.
\end{align}
The ordinary second-order scattering [i.e., the first line of Eq.~\eqref{fdavfvd}] induces decoherence—manifest as the decay of off-diagonal elements of the density matrix—which is characterized by a normal decoherence rate:
\begin{align} \label{bgsfgkf}
    \Gamma_{k}/\hbar=\frac{1}{8}\left(1+\frac{\epsilon^2_L}{\mathcal{E}^2_{k}}\right)\left(\frac{1}{\tau^{0}_{k+}}+\frac{1}{\tau^{0}_{k-}}\right),
\end{align}
where $\frac{1}{\tau^0_{k\eta}}=(2\pi/\hbar)n_{\text{i}}\nu_{\eta}(\epsilon_{k\eta})U^{2}$. More importantly, the anomalous scattering [i.e., the second line of Eq.~\eqref{fdavfvd}], characterized by an anomalous decoherence rate 
\begin{align} \label{bgsfgkfa}
    \Gamma^a_{k}/\hbar= \frac{\epsilon_L}{4\mathcal{E}_{k}}  \left(\frac{1}{\tau^{0}_{k+}}+\frac{1}{\tau^{0}_{k-}}\right),
\end{align}
serves as a transverse drive field, thereby behaving as an effective out-of-plane magnetic field.

The off-diagonal component of the quantum kinetic equation \eqref{fvdvkfvkmain1} becomes 
\begin{align} \label{vfkfvkmd}
    \frac{\partial}{\partial t}  \delta\varrho^{\bar{\eta}\eta}_{\vec{k}}&+\frac{i}{\hbar}(\epsilon_{k\bar{\eta}}-\epsilon_{k\eta}) \delta\varrho^{\bar{\eta}\eta}_{\vec{k}}+e\frac{i}{\hbar}E^i\mathcal{R}^{i,\bar{\eta}\eta}_{\vec{k}} (f_{k\bar{\eta}}- f_{k\eta}) \notag \\
    &=-\frac{1}{\hbar}\Gamma_{k} \delta\varrho^{\bar{\eta}\eta}_{\vec{k}}+i\eta\frac{1}{\hbar}\Gamma^a_{k}\delta\varrho^{\bar{\eta}\eta}_{\vec{k},a}.
\end{align}
Using Eqs.~\eqref{ansatz} and \eqref{gbbkgbm}, Eq.~\eqref{vfkfvkmd}, in steady state ($ \partial_t  \delta\varrho^{\bar{\eta}\eta}_{\vec{k}}$=0), is separated into ordinary and anomalous components of density matrix, i.e., $\delta\varrho^{\bar{\eta}\eta}_{\vec{k},\Vert}$ and $\delta\varrho^{\bar{\eta}\eta}_{\vec{k},\perp}$, 
\begin{align} \label{ovfkfvkmd}
    &+\frac{i}{\hbar}(\epsilon_{k\bar{\eta}}-\epsilon_{k\eta})\delta\varrho^{\bar{\eta}\eta}_{\vec{k},\Vert} +e\frac{i}{\hbar}E^i\mathcal{R}^{i,\bar{\eta}\eta}_{\vec{k}} (f_{k\bar{\eta}}- f_{k\eta}) \notag \\
    &=-\frac{\Gamma_{k}}{\hbar} \delta\varrho^{\bar{\eta}\eta}_{\vec{k},\Vert}+i\eta\frac{\Gamma^a_{k}}{\hbar}\left(-\frac{\tau^{\bar{\eta}\eta}_{k,\perp}}{\tau^{\bar{\eta}\eta}_{k,\Vert}}\right)\delta\varrho^{\bar{\eta}\eta}_{\vec{k},\Vert},
\end{align}
\begin{align} \label{avfkfvkmd}
    +\frac{i}{\hbar}(\epsilon_{k\bar{\eta}}-\epsilon_{k\eta})\delta\varrho^{\bar{\eta}\eta}_{\vec{k},\perp} =-\frac{\Gamma_{k}}{\hbar} \delta\varrho^{\bar{\eta}\eta}_{\vec{k},\perp}+i\eta\frac{\Gamma^a_{k}}{\hbar}\frac{\tau^{\bar{\eta}\eta}_{k,\Vert}}{\tau^{\bar{\eta}\eta}_{k,\perp}}\delta\varrho^{\bar{\eta}\eta}_{\vec{k},\perp}.
\end{align}
Together with Eq.~\eqref{fvfkvmkdf4} and by solving the system of equations above, we obtain the following solutions--decoherence time--as follows  
\begin{align} \label{sgfbf10}
    \frac{\tau^{\bar{\eta}\eta}_{k,\Vert}}{\hbar}=\frac{(\epsilon_{k\bar{\eta}}-\epsilon_{k\eta})-i\Gamma_{k}}{[(\epsilon_{k\bar{\eta}}-\epsilon_{k\eta})-i\Gamma_{k}]^2+(\Gamma^a_{k})^2},
\end{align}
\begin{align} \label{sgfbf20}
    \frac{\tau^{\bar{\eta}\eta}_{k,\perp}}{\hbar}=\frac{\eta\Gamma^a_{k}}{[(\epsilon_{k\bar{\eta}}-\epsilon_{k\eta})-i\Gamma_{k}]^2+(\Gamma^a_{k})^2}.
\end{align}

\textit{Appendix B: Comparison of different extrinsic mechanisms--}  We next analyze the Hall conductivities arising from different extrinsic mechanisms. At zero temperature ($T=0$), our AHE conductivities are given by Eqs.~\eqref{oHov} and \eqref{oHop},  while those of  side jump and skew scattering are given by~\cite{sinitsyn2007anomalous,zhang2025theory} 
\begin{align}\label{oHsj}
    \sigma_{H}^{\text{sj}}=\frac{2e^2}{h}\sum_{\eta}\eta \frac{\epsilon_L(v_Rk^{\eta}_F)^2}{\mathcal{E}_{k^{\eta}_F}\left[4\epsilon^2_L+(v_Rk^{\eta}_F)^2\right]},
\end{align}
\begin{align}\label{oHsk}
    \sigma_{H}^{\text{sk}}=-\frac{4e^2}{h}\frac{}{}\sum_{\eta}\eta\frac{\epsilon_L (k^{\eta}_F)^2U\mathcal{E}^3_{k^{\eta}_F}(k^{\eta}_Fl_{\text{so}})^2}{\left[4\epsilon^2_L+(v_Rk^{\eta}_F)^2\right]^2}\frac{\tau^{0}_{k^{\eta}_F\eta}}{\hbar}.
\end{align}
Here $l_{\text{so}}=\frac{\hbar}{2mc}$ and $c$ is speed of light.  Figures~\ref{FIG1N} (a) and (b) show the $n_{\text{i}}$ dependence of various extrinsic contributions for different $\epsilon_{L}$. The side-jump contribution (blue curves) is independent of $n_{\mathrm{i}}$~\cite{sinova2015spin,chang2023colloquium} and can therefore be readily isolated experimentally. Next, we compare our mechanism with skew scattering one, which is expected to be linear in longitudinal conductivity, $\sigma_{xx}\propto 1/n_{\mathrm{i}}$ [inset of Fig.~\ref{FIG1N} (a)]. For small $n_{\text{i}}$, the impurity-density scaling of our decoherence-induced mechanism ($\delta\sigma^{\Vert}_{H}, \sigma^{\perp}_{H} \propto n^2_{\mathrm{i}}$) is fundamentally different from that of skew scattering ($\sigma_{H}^{\text{sk}}\propto 1/n_{\mathrm{i}}$). Furthermore, our Hall conductivity—originating from a second-order scattering process intimately associated with decoherence—is significantly larger than the  skew-scattering contribution, as illustrated by the green curves in Fig.~\ref{FIG1N}. Comparing longitudinal conductivity $\sigma_{xx}=\frac{e^2}{2h}\sum_{\eta} \frac{4\mathcal{E}^2_{k^{\eta}_F}}{4\epsilon^2_L+(v_Rk^{\eta}_F)^2}k^{\eta}_F   v_{k^{\eta}_F\eta} \tau^{0}_{k^{\eta}_F\eta}$ and $\sigma_{H}^{\text{sk}}$, the conversion efficiency for $\eta$ band electrons is $\theta^{\text{sk}}_{\eta}=\eta\pi\frac{4\epsilon_L\mathcal{E}_{k^{\eta}_F}}{4\epsilon^2_L+(v_Rk^{\eta}_F)^2}(k^{\eta}_Fl_{\text{so}})^2\nu_{\eta}(\epsilon_{k^{\eta}_F\eta})U$. In Fig.~\ref{FIG1N}, $(k^{\eta}_Fl_{\text{so}})^2$ and $\nu_{\eta}(\epsilon_{k^{\eta}_F\eta})U$ are $\sim 10^{-8}$ and $\sim 10^{-1}$, respectively. Consequently, an appreciable magnetic-field modulation of the AHE that exhibits a strong quadratic-in-$n_{\text{i}}$ scaling provides a clear experimental signature of the extrinsic mechanism associated with decoherence.

\textit{Appendix C: $T$ dependence of our AHE--} At nonzero temperature, 
the Hall conductivity from  $\delta\varrho^{\bar{\eta}\eta}_{\vec{k},\Vert}$ and  $\delta\varrho^{\bar{\eta}\eta}_{\vec{k},\perp}$, are given as follows:
\begin{align}  \label{temHov}
    \sigma^{\Vert}_{H}(T)&=-\frac{e^2}{h}\frac{v_R }{\hbar}\int^{\infty}_{0}dk\frac{v_Rk}{\mathcal{E}_k}\frac{\epsilon_L}{\mathcal{E}_k}\text{Re} (\tau^{+-}_{k,\Vert})\notag\\
     &\times\left[\frac{1}{(e^{\beta(\epsilon_{k+}-\epsilon_{F})} + 1)}-\frac{1}{(e^{\beta(\epsilon_{k-}-\epsilon_{F})} + 1)}\right],
\end{align}
\begin{align} \label{temgbllgla4}
    \sigma^{\perp}_{H}(T)&=\frac{e^2}{ h} \frac{v_R}{2\hbar} \int^{\infty}_{0} dk\left(1+\frac{\epsilon^2_L}{\mathcal{E}^2_k}\right)\frac{v_Rk}{\mathcal{E}_k}\text{Im} (\tau^{+-}_{k,\perp})\notag\\
    &\times\left[\frac{1}{(e^{\beta(\epsilon_{k+}-\epsilon_{F})} + 1)}-\frac{1}{(e^{\beta(\epsilon_{k-}-\epsilon_{F})} + 1)}\right].
\end{align}
Their $T$ dependence  are plotted in Fig. \ref{FIG2} for different values of $n_{\text{i}}$. The AHE conductivity from skew scattering, linear in $\sigma_{xx}(T)$, is expected to be almost constant for temperature lower than the Debye temperature $T_D$~(e.g. $T_D=470$ K for Fe) where the $T$ modulation via the electron-phonon interaction is omitted. 
However, in our decoherence theory, the non-equilibrium coherence \eqref{fvfkvmkdf4} strongly depends on $T$, resulting in a substantially strong modulation of the our decoherence-relate AHE on $T$. When $\epsilon_F>\epsilon_L$, such that the Fermi energy intersects both $\eta=-$ and $\eta=+$ bands, the conductivity exhibits a maximum as a function of $T$, reminiscent of the Kondo effect. This feature disappears when $\epsilon_F<\epsilon_L$, where the Fermi energy intersects only the $\eta=-$ band.

\end{document}